\documentclass[aps,prl,reprint,twocolumn,superscriptaddress]{revtex4-1}

\usepackage{amsmath,amsthm,amsfonts,amssymb,bm}
\usepackage{times}
\usepackage[colorlinks={true},dvipdfm]{hyperref}
\hypersetup{citecolor={blue}, filecolor={blue}, linkcolor={blue}, urlcolor={blue}}

\begin{document}

\noindent {\large\bf Comment on ``Are There Traps in Quantum Control Landscapes?''}
\vskip 0.12in

Many quantum control problems are formulated as a search for an optimal field that maximizes a physical objective. This search is performed over a landscape defined as the objective as a function of the control field. A recent Letter \cite{PechenTannor2011} asserts that the existence of special landscape critical points (CPs) with trapping characteristics is ``contrary to recent claims in the literature'' and ``can have profound implications for both theoretical and experimental quantum control studies.'' We show here that these assertions are inaccurate and misleading.

First, the authors in \cite{PechenTannor2011} declare that their finding of local traps at singular CPs corresponding to constant control fields contradicts the existing theory of quantum control landscapes. However, in making this statement, they completely focus on an early paper published in 2004 \cite{Rabitz2004} and improperly ignore the large body of subsequent research, in which a careful distinction has been made between regular and singular CPs (see Refs.~\cite{Chakrabarti2007, Brif2010} for reviews). Specifically, the current status of the theory is that the equivalence of kinematic and dynamic control landscapes and the resulting absence of local traps applies only to regular CPs (at which the tangent map from the space of controls to the space of evolution operators is locally surjective) \cite{Brif2010}. No such assessment is made in the literature for singular CPs (including constant-field solutions reported in \cite{PechenTannor2011}). Already in 2007, Chakrabarti and Rabitz \cite{Chakrabarti2007} discussed the existence of singular CPs at constant control fields and clearly stated that the results regarding the trap-free character of control landscapes apply only to regular CPs. This distinction between regular and singular CPs and the applicability of the trap-free result only to regular ones was further stated in a number of subsequent works \cite{Brif2010, LandscapeRegCP}. With regard to singular CPs, it was only assumed that their measure in the search space is much smaller than that of regular ones, and this assumption is fully supported by the literature (see below). Correspondingly, no specific assessment is made in the current landscape theory with regard to the optimality of singular CPs, hence the report in \cite{PechenTannor2011} of locally optimal singular controls in no way contradicts this theory.

Second, the statement in \cite{PechenTannor2011} with regard to implications of their work for quantum control studies is not supported by any evidence. In fact, the singular CPs reported in \cite{PechenTannor2011} are special cases of constant (e.g., zero) fields with strictly limited systems drawn from a null set of controls and Hamiltonians; moreover, additional requirements are imposed on the target observable, which is forced to have a particular form, rather than being determined by physical considerations. The controls, Hamiltonians, and observables satisfying these unrealistically demanding conditions form a very restricted set, and thus do not have general physical significance. What matters in practice is not that traps may exist mathematically under such specially tailored conditions, but rather their likelihood of being encountered under broad physical circumstances. To test this matter, the recent literature contains two studies \cite{Moore2011} that involve many thousands of optimization runs for state-transition and evolution-operator control with a variety of systems. Contrary to the assertion in \cite{PechenTannor2011}, none of these simulations encountered traps and all achieved maximum objective values upon due care to numerical details. In reported cases where optimization runs are trapped, this happens due to excessive constraints on controls. For example, in Ref.~\cite{FouquieresSchirmer2010}, the trapped searches were caused by the control time $T$ being too short (see Ref.~\cite{MooreBrif2011} for explanation of this effect); this trapping has nothing to do with fundamental properties of the control landscape and is easily eliminated simply by increasing $T$.

While it is well known \cite{Chakrabarti2007, Brif2010} that singular CPs may exist, there is no evidence that they pose any obstacle to searches for globally optimal controls. For example, a recent numerical study \cite{WuDominy2009} designed to identify singular CPs found none that are traps. To further investigate this issue, we performed several thousand quantum control simulations \cite{note} for the special classes of systems and target observables proposed in \cite{PechenTannor2011}. In the case of kinematic CPs in a $\Lambda$-type system \cite{PechenTannor2011}, our simulations show that the presence of a second-order trap at zero field has no effect on convergence to a globally optimal solution unless the initial field is intentionally made very close to zero. In the case of nonkinematic CPs \cite{PechenTannor2011}, we encountered no trapping whatsoever in the vicinity of a CP at zero field. These results are in full agreement with the general observation that local optima are not found to have any impact in a wide range of applications. The existing landscape theory \cite{Chakrabarti2007, Brif2010} provides the basic foundation to understand this observation.
 
\vskip 0.1in

\noindent {Herschel Rabitz,$^1$ Tak-San Ho,$^1$ Ruixing Long,$^1$
Rebing Wu,$^2$ and Constantin Brif$^3$}\\
\hspace*{0.1in} 
\parbox[t]{3.2in}{\small{$^1$Department of Chemistry, Princeton
University \\ Princeton, New Jersey 08544, USA \\ 
$^2$Department of Automation, Tsinghua University \\ 
Beijing 100084, P.R. China \\
$^3$Sandia National Laboratories, Livermore \\
California 94550, USA}}


\vskip -0.15in


\end{document}